\RequirePackage{fix-cm}
\documentclass[twocolumn,final]{svjour3}     
\smartqed  
\usepackage{graphicx}
\usepackage{amsmath}
\usepackage{amssymb}
\usepackage{psfrag}
\usepackage{color}
\usepackage{stfloats}
\usepackage{mathrsfs}
\usepackage{tabularx}
\usepackage{booktabs}
\usepackage{colortbl}
\usepackage{rotating}
\usepackage{boldline}
\usepackage{bm}
\usepackage{changes}
\usepackage{url}
\usepackage{citesort}
\usepackage{adjustbox}
\usepackage{graphicx}
\usepackage{cellspace}
\DeclareMathOperator*{\argmin}{arg\,min}
\usepackage{lineno}

\journalname{}
\newcommand{\beq}{\begin{equation}}
\newcommand{\eeq}{\end{equation}}

\def\vecs  #1{{\rm{\bf{#1}}}{}}
\let\vec\bm
\let\ten\bm
\begin{document}
\title{Discovering uncertainty: \\
Gaussian constitutive neural networks with correlated weights}
\author{Jeremy A. McCulloch \and Ellen Kuhl}
\institute{Jeremy A. McCulloch \at
              318 Campus Drive, Stanford, CA, 94305 \\
              \email{jmcc@stanford.edu}           
\and 
Ellen Kuhl \at
              318 Campus Drive, Stanford, CA, 94305 \\
              \email{ekuhl@stanford.edu}           
}

\date{Received: March 16, 2025}
\maketitle

\begin{abstract} %
When characterizing materials, it can be important to not only predict their mechanical properties, but also to estimate the probability distribution of these properties across a set of samples. Constitutive neural networks allow for the automated discovery of constitutive models that exactly satisfy physical laws given experimental testing data, but are only capable of predicting the mean stress response. Stochastic methods treat each weight as a random variable and are capable of learning their probability distributions. Bayesian constitutive neural networks combine both methods, but their weights lack physical interpretability and we must sample each weight from a probability distribution to train or evaluate the model. Here we introduce a more interpretable network with fewer parameters, simpler training, and the potential to discover correlated weights: Gaussian constitutive neural networks. We demonstrate the performance of our new Gaussian network on biaxial testing data, and discover a sparse and interpretable four-term model with correlated weights. Importantly, the discovered distributions of material parameters across a set of samples can serve as priors to discover better constitutive models for new samples with limited data. We anticipate that Gaussian constitutive neural networks are a natural first step towards generative constitutive models informed by physical laws and parameter uncertainty. \\[4.pt]
Our source code, data, and examples are available at \\https://github.com/LivingMatterLab/CANN.
\keywords{Gaussian neural networks \and
uncertainty quantification \and
constitutive artificial neural networks \and
constitutive modeling \and
automated model discovery}
\end{abstract}
\section{Motivation}
Standard constitutive models predict the stress in a material from its deformation \cite{holzapfel_2000}, but do not provide error bars for this estimate or characterize the probability distribution of the stress. Knowing the uncertainty of a model prediction is particularly important for biological and surgical materials, which are used in finite element simulations to predict outcomes of a surgical procedure or disease progression \cite{updegrove_simvascular_2017,fletcher_development_2014}. The accuracy of these finite element simulations is intrinsically limited by the accuracy of the underlying constitutive model \cite{peirlinck_2024}. Furthermore, experimental data for biological materials often display large error bars \cite{budday_mechanical_2017,sommer_biomechanical_2015}, mainly caused by the inherent 
inter-patient variability, but also by varying boundary effects from inconsistent experimental settings \cite{miller_method_2005}. Existing constitutive machine learning methods such as constitutive artificial neural networks \cite{linka_new_2023}, EUCLID \cite{flaschel_automated_2023},  and neural ordinary differential equations \cite{tac_data-driven_2024} can discover constitutive models that accurately predict the {\it{mean stress response}}, but typically do not quantify the {\it{variance of the stress}} \cite{fuhg_2024}. Recent studies have discovered constitutive models for the average across several samples compared to each individual sample, and emphasize need for a unified model that takes uncertainty into account \cite{vervenne_constitutive_2025}. To use constitutive models with confidence, we need interpretable and validated models that can not only accurately predict the mean stress response but also characterize the variance of the stress. 

\subsection{Constitutive Modeling}
A hyperelastic constitutive model is a function that predicts the local stress in a material given its local deformation. The global deformation of a material is defined by its deformation map $\vec x = \ten{\varphi}(\vec X)$ that maps a point $\vec X$ in the reference configuration to a point $\vec x$ in the deformed configuration. The local deformation is measured by the deformation gradient $\ten{F} = \nabla_{\vecs X} \ten{\varphi}(\vec X)$ and the right Cauchy-Green deformation tensor $\ten{C} = \ten{F}^{\rm{t}} \cdot \ten{F}$. The local stress in a material is defined by its Piola stress $\ten{P}$. While a generic Cauchy-elastic constitutive model can express the Piola stress as $\ten{P} = \ten f(\ten{F})$ for any function $\ten f$, hyperelastic constitutive models restrict the Piola stress to the derivative of a scalar potential $\ten{P} = \partial  \psi(\ten{F}) / \partial \ten{F}$. Various approaches have been used to model the strain energy function $\psi$, including classical continuum-mechanics-based approaches and machine-learning-based approaches such as constitutive neural networks. 

\subsection{Constitutive Neural Networks}
Constitutive neural networks combine classical constitutive modeling approaches with modern machine learning techniques to automatically discover interpretable constitutive models that satisfy physical constraints \cite{linka_new_2023}. The classical method of discovering a constitutive model consists of selecting a model from a library of phenomenological models and then optimizing the parameters of that model. This process requires domain expertise, is iterative, and is potentially time-consuming \cite{linka_new_2023}. 
Neural networks offer a solution that can approximate any nonlinear function \cite{cybenko_approximation_1989} given sufficient training data and minimal human input; however, standard fully-connected neural networks fail to extrapolate to unseen loading conditions, and can violate known physical constraints such as polyconvexity or material symmetry \cite{buganza_2025}. Physics-informed neural networks offer a method of incorporating physical constraints into neural networks by introducing an additional term into the loss function \cite{raissi_physics-informed_2019}. However, physics-informed neural networks do not exactly satisfy physical laws unless the physics-informed loss is zero. Constitutive neural networks use carefully chosen activation functions and model architectures to ensure that physical constraints are satisfied by construction, regardless of the model weights. These physical constraints include polyconvexity, material objectivity, and symmetry. Different constitutive neural network architectures are used depending on the type of material, with specific models developed for isotropic materials \cite{pierre_principal-stretch-based_2023}, transversely isotropic materials \cite{vervenne_constitutive_2025}, and inelastic materials \cite{holthusen_theory_2024,boes_accounting_2024}; for orthotropic meshes, a model has been developed that has eight isotropic terms and six anisotropic terms \cite{mcculloch_automated_2024}. 

This orthotropic constitutive neural network represents the strain energy as a function of the first and second invariants, $I_1 = \ten{C}:\ten{I}$ and $I_2 = \frac{1}{2} [\,I_1^2 - \ten{C}:\ten{C}\,]$, as well as three mixed invariants, $I_{4w} = \vec{w} \cdot \ten{C} \cdot \vec{w}$, $I_{4s_I} = \vec{s}_I \cdot \ten{C} \cdot \vec{s}_I$, and $I_{4s_{II}} = \vec{s}_{II} \cdot \ten{C} \cdot \vec{s}_{II}$. The unit vector $\vec{w}$ is the direction of the stiffest fiber, and the unit vectors $\vec{s}_{I}$ and $\vec{s}_{II}$ are offset from $\vec{w}$ by 60 degrees in each direction. These invariants are raised to the first and second powers and either identity or exponential activation functions are applied. The complete strain energy function in terms of the internal and external weights 
$w_i$ and $w_i^*$ takes the following form,
\begin{equation*}
\begin{array}{l@{\hspace*{.10cm}}c@{\hspace*{.10cm}}l@{\hspace*{.04cm}}
              l@{\hspace*{.04cm}}l@{\hspace*{.04cm}}l@{\hspace*{.00cm}} 
              l@{\hspace*{.04cm}}l@{\hspace*{.04cm}}l@{\hspace*{.04cm}}
              l@{\hspace*{.04cm}}l@{\hspace*{.00cm}}l}
    \psi 
&=& w_{1} &w_{1}^* &[ I_1 &- 3 ]
&+& w_{2} & [ \, \exp ( w_{2}^* & [ I_1 &-3 ]&)   - 1] \\
&+& w_{3} &w_{3}^* &[ I_1 &- 3 ]^2
&+& w_{4} & [ \, \exp ( w_{4}^* & [ I_1 &-3 ]^2&) - 1] \\
&+& w_{5} &w_{5}^* &[ I_2 &- 3 ]
&+& w_{6} & [ \, \exp ( w_{6}^* & [ I_2 &-3 ]&)   - 1] \\
&+& w_{7} &w_{7}^* &[ I_2 &- 3 ]^2
&+& w_{8} & [ \, \exp ( w_{8}^* & [ I_2 &-3 ]^2&) - 1] \\
&-& w_{9} &w_{9}^* &[ I_{4w} &- 1 ]
&+& w_{9} & [ \, \exp ( w_{9}^* & [ I_{4w} &-1 ]&)   - 1] \\ 
&+& w_{10} &w_{10}^* &[ I_{4w} &- 1 ]^2
&+& w_{11} & [ \, \exp ( w_{11}^* & [ I_{4w} &-1 ]^2&) - 1] \\
&-& w_{12} &w_{12}^* &[ I_{4s} &- 1 ]
&+& w_{12} & [ \, \exp ( w_{12}^* & [ I_{4s} &-1 ]&)   - 1] \\
&+& w_{13} &w_{13}^* &[ I_{4s} &- 1 ]^2
&+& w_{14} & [ \, \exp ( w_{14}^* & [ I_{4s} &-1 ]^2&) - 1] \,.
\label{CANNenergy_2fibers}
\end{array}
\end{equation*}
By using the chain rule, we can obtain an analytical expression for the Piola stress, $\ten{P} = \partial  \psi(\ten{F}) / \partial \ten{F}$, in terms of the deformation gradient $\ten{F}$ and the weights $w_i$ and $w_i^*$. 
In a perfectly homogeneous and shear-free biaxial extension test
the deformation gradient 
$\ten{F} = {\rm{diag}} \, \{\,\lambda_1,\lambda_2,(\lambda_1 \lambda_2)^{-1}\,\}$
and the Piola stress
$\ten{P} = {\rm{diag}} \, \{\,P_{11},P_{22},0\,\}$,
remain diagonal, 
where $\lambda_1$ and $\lambda_2$ are the biaxial stretches.
The only non-zero stress components are the two normal stresses $P_{11}$ and $P_{22}$ in the two loading directions, 
$$
P_{11} \!= \!\mbox{$\sum_{i=1}^{N}$} w_i  f_i (\ten{F}; w_i^*) -p
\; \mbox{and} \;
P_{22} \!= \!\mbox{$\sum_{i=1}^{N}$} g_i(\ten{F}; w_i^*) -p,
$$
where $p=2 (\lambda_1 \lambda_2)^{-2} \, \partial \psi/\partial I_1 + 2 (\lambda_1^{-2}+\lambda_2^{-2}) \, \partial \psi/\partial I_2$ is the pressure term to satisfy the incompressibility condition,
$N=14$ is the number of network terms, and
$$
f_i\,(\ten{F}; w_i^*) = \frac{1}{w_i} \, \frac{\partial \psi_i}{\partial F_{11}}
\quad \mbox{and} \quad 
g_i\,(\ten{F}; w_i^*) = \frac{1}{w_i} \, \frac{\partial \psi_i}{\partial F_{22}}
$$
are functions of only $F_{11}$ or $F_{22}$ and of the internal weights $w_i^*$. To train a constitutive  neural network, we typically use a numerical optimizer to find the weights $w_i$ and $w_i^*$ that minimize the loss function. A typical loss function is the mean squared error between the model predicted stress and the measured stress. This loss function is sometimes weighted to put more emphasis on some experimental data, and is often combined with $L_p$ regularization to encourage sparsity and avoid overfitting \cite{mcculloch_sparse_2024,hou_2025}. 
\subsection{Gaussian Neural Networks}
In a standard artificial neural network, the model architecture defines a function $\phi$, which maps the input $x_i$ and weights $\vec{w}$ to a predicted output $y_i$ as $y_i = \phi(x_i; \vec{w})$. 
Next, a numerical optimization is performed to find the weights $\bar{\vec{w}}$ that minimize the mean squared error between the model predictions $y_i$ and the data $\hat{y}_i$,
such that $\bar{\vec{w}} = \argmin_{\vecs{w}} \sum_{i=1}^M (\phi(x_i; \vec{w})-\hat{y}_i)^2$
where $M$ is the number of data points.
In a probabilistic framework, this optimization can be viewed as finding a maximum likelihood estimate of $\vec{w}$ when we model $\hat{y}_i = y_i + \epsilon_i$, where $\epsilon_i 
{\sim} {\mathcal{N}}(0, \sigma^2)$ for any constant $\sigma$. Under this assumption, we can write the negative log likelihood as 
$$
{\textsf{NLL}} = -\ln\left(\prod_{i=1}^M \frac{1}{\sqrt{2\pi \sigma^2}} \exp\left(-\frac{(y_i  - \hat{y}_i)^2}{2\sigma^2}\right)\right)\,,
$$
which we can reformulate as follows,
$$
{\textsf{NLL}} = \frac{1}{2} \,M \ln(2\pi \sigma^2) + \frac{1}{2\sigma^2}\sum_{i=1}^M (y_i - \hat{y}_i)^2\,.
$$
Thus, the negative log likelihood is an increasing affine function of the mean squared error, so minimizing the mean squared error and minimizing the negative log likelihood are equivalent. 

In a Gaussian neural network \cite{nix_estimating_1994}, the neural network predicts not only the expected value of the output $y_i$, but also the variance of the noise $\sigma^2$. There are various ways to model the mean and variance, but all models consist of functions $\mu(x_i; \vec{w})$ and $\sigma^2(x_i; \vec{w})$ that map the input $x_i$ and weights $\vec{w}$ to the model mean and variance. Next, we can find a maximum likelihood estimate of the weights $\vec{w}$ when the output is distributed according to the distribution, $y_i \sim {\mathcal{N}}(\mu(x_i; \vec{w}), \sigma^2 (x_i; \vec{w}))$. To find the maximum likelihood estimate, we can compute the negative log likelihood as a function of $\mu(x_i; \vec{w})$ and $\sigma^2(x_i; \vec{w})$ and use gradient descent methods to find the weights $\bar{\vec{w}}$ that minimize this quantity. Importantly, because the variance $\sigma^2$ is no longer a constant, this optimization is no longer equivalent to minimizing the mean squared error. 

\subsection{Bayesian Neural Networks}
As with a Gaussian neural network, a Bayesian neural network specifies a probability distribution for the output as a function of the model inputs and weights. This is often a Gaussian distribution, $y_i \sim {\mathcal{N}}(\mu(x_i; \vec{w}), \sigma^2(x_i; \vec{w}))$ where $\sigma^2$ is either a function of the input or a constant. However, in a Bayesian network, the weights $\vec{w}$ are treated as random variables with known priors ${\textsf{P}}(\vec{w})$. Instead of performing maximum likelihood estimation on the weights, we calculate the posterior ${\textsf{P}}(\vec{w}|\hat{D})$ of the weights $\vec{w}$ for given data $\hat{D}$,
$$
{\textsf{P}}(\vec{w}|\hat{D})
= \frac{{\textsf{P}}(\vec{w}) \, {\textsf{P}}(\hat{D}|\vec{w})}{{\textsf{P}}(\hat{D})}
$$
by multiplying the likelihood ${\textsf{P}}(\hat{D}|\vec{w})$ by the prior ${\textsf{P}}(\vec{w})$ divided by the unknown evidence ${\textsf{P}}(\hat{D})$. To approximate this intractable posterior, we define a tractable distribution $\textsf{Q} (\vec{w}; \theta)$ parameterized by $\theta$. Then, to train the Bayesian neural network, we calculate the parameters $\bar{\theta}$ that minimize the KL divergence between $\textsf{Q} (\vec{w}; \theta)$ and the posterior $\textsf{P}(\vec{w}|\hat{D})$. To evaluate a Bayesian neural network on some input $x$, we first sample the weights $\vec{w}$ from the tractable distribution $\textsf{Q} (\vec{w}; \bar{\theta})$ and then evaluate the neural network using the sampled weights. 

While both Gaussian and Bayesian neural networks incorporate uncertainty, the principal difference is that priors for the weights are specified in Bayesian neural networks. This implies that the parameters identified using Bayesian neural networks approximate the posteriors of the weights conditioned on the training data. By contrast, the parameters identified using Gaussian neural networks are maximum likelihood estimates of the weights, and are learned entirely from the data, without having to choose a prior. 

In a fully connected neural network, we expect the weights to be distributed according to a Gaussian distribution with zero mean and a small variance that depends on the number of nodes per layer \cite{glorot_understanding_2010}. Consequently, the priors for Bayesian networks are often a Gaussian distribution with zero mean, or a spike-and-slab prior if sparse networks are desired \cite{goan_bayesian_2020}. However, in constitutive neural networks, we expect the weights to be non-negative, and the expected magnitude of the weights depends on the material stiffness. Thus, without any domain knowledge, selecting appropriate priors for constitutive neural networks is much more challenging. 

\subsection{Bayesian Constitutive Neural Networks}
Previous work has combined constitutive neural networks with Bayesian neural networks to create a model that incorporates both physical constraints and uncertainty \cite{linka_discovering_2025}. Each stress term in this model is computed similarly to a deterministic constitutive neural network in terms of the external and internal weights $w_i$ and $w_i^*$,
$$
P_{11,i} = w_i \, f_i(\ten{F}; w_i^*) -p
\quad \mbox{and} \quad  
P_{22,i} = w_i \, g_i(\ten{F}; w_i^*)-p .
$$
Then, to obtain the means and standard deviations, we would 
sum all stress terms weighted by their individual probability densities 
${\mathcal{N}}(w_{\mu,i}, w_{\sigma,i})$
in terms of the  variational weights $w_{\mu,i}$ and $w_{\sigma,i}$,
\[
\begin{array}{@{\hspace*{1.8cm}}l}
P_{11} = \!\mbox{$\sum_{i=1}^N$} {\mathcal{N}}(w_{\mu,i}, w_{\sigma,i}) \, P_{11,i} \\  [2.pt]
P_{22} = \!\mbox{$\sum_{i=1}^N$} {\mathcal{N}}(w_{\mu,i}, w_{\sigma,i}) \, P_{22,i}.
\end{array}
\] 
This approach introduces four sets of weights, the deterministic external and internal network weights $w_i$ and $w_i^*$, and the variational weights, $w_{\mu,i}$ and $w_{\sigma,i}$. Since the variational weights are parameterized in terms of a Gaussian distribution, each contributes two trainable parameters, so the total number of trainable parameters per term is six. The network learns these parameters by maximizing the evidence lower bound. \\[2.pt]

The goal of this work is to develop a {\it{Gaussian constitutive neural network}} that is more interpretable than the existing Bayesian constitutive neural network and {\it{does not require prior selection}}. We reduce the number of trainable parameters per term from six to three, which {\it{decreases model complexity}} and {\it{improves model interpretability}}. In addition, we perform maximum likelihood estimation instead of maximizing the evidence lower bound which eliminates the need to select prior distributions for the weights. Instead of drawing the variational weights from their learnt distributions to calculate the stress, we compute the model distribution of the stress using probability theory. Finally, we allow for {\it{correlated  weights}} instead of forcing the weights of the individual terms to be independent. We demonstrate that these modifications results in improved model performance during both training and testing. 

\section{Methods}\label{sec_kinematics}
\subsection{Model Architecture}\label{sec:arch}
Similar to the previous sections, we can write the stress for an orthotropic constitutive neural network as 
$$
P_{11} = \mbox{$\sum_{i=1}^N$} \, w_i f_i(\ten{F}; w_i^*)
\;\, \mbox{and} \;\,
P_{22} = \mbox{$\sum_{i=1}^N$} \, w_i g_i(\ten{F}; w_i^*)\,,
$$
where $\ten{F}$ is the deformation gradient corresponding to the $j$-th training set, 
$w_i$ and $w_i^*$ are the external and internal network weights, 
and 
$$
f_i\,(\ten{F}; w_i^*) = \frac{1}{w_i} \, \frac{\partial \psi_i}{\partial F_{11}}
\quad \mbox{and} \quad 
g_i\,(\ten{F}; w_i^*) = \frac{1}{w_i} \, \frac{\partial \psi_i}{\partial F_{22}}
$$
are functions that are defined by the network model. In a deterministic constitutive neural network, we learn values of $w_i$ and $w_i^*$ that minimize the mean squared error between the data and the model prediction. In a Gaussian constitutive neural network, we let
$w_i \sim {\mathcal{N}}(w_{\mu,i}, (w_{i,\mu} w_{i,\sigma})^2)$ where 
${\mathcal{N}}$ denotes the normal distribution with mean $w_{i,\mu}$ and variance $(w_{i,\mu} w_{i,\sigma})^2$, and then learn the values of $w_{\mu,i}$ and $w_{\sigma,i}$ and $w_i^*$ that maximize the likelihood of observing the experimental data under the model. In contrast to previous constitutive neural networks with independently distributed external weights $w_i$, 
we allow the weights $w_i$ to be correlated with one another. We summarize the stress contributions of the individual nodes in the vectors 
\[
\begin{array}{@{\hspace*{0.8cm}}l}
\hat{\vec{P}}_{11} 
= [w_{\mu,1} f_1(\ten{F}; w_1^*), \ldots, w_{\mu,N} f_N(\ten{F}; w_N^*)] \\[4.pt]
\hat{\vec{P}}_{22} 
= [w_{\mu,1} g_1(\ten{F}; w_1^*), \ldots, w_{\mu,N} g_N(\ten{F}; w_N^*)],
\end{array}
\] 
and express the joint distribution of the weights $w_i$ as
$$
\hat{\vec{w}}
= \left[ \frac{w_1}{w_{\mu,1}}, \frac{w_2}{w_{\mu,2}}, \ldots, \frac{w_N}{w_{\mu,N}} \right] \sim {\mathcal{N}}([1, 1, \ldots , 1], \ten{\varSigma})$$
for any positive semidefinite $\ten{\varSigma}$. 
This allows us to rewrite the stresses $P_{11}$ and $P_{22}$ as vector products,
$$
P_{11} = \hat{\vec{P}}_{11} \cdot \hat{\vec{w}}
\quad \mbox{and} \quad
P_{22} = \hat{\vec{P}}_{22} \cdot \hat{\vec{w}} \,.
$$
Since $\hat{\vec{P}}_{11}$ and $\hat{\vec{P}}_{22}$ are constant and $\hat{\vec{w}}$ is a random variable distributed according to a multivariate normal distribution, the resulting normal stresses $P_{11}$ and $P_{22}$ are random variables with normal distributions. The means $\mu_{11}$ and $\mu_{22}$ of these normal distributions are
\[
\begin{array}{@{\hspace*{0.1cm}}l}
  \mu_{11}
= {\textsf{E}}[P_{11}] 
= \hat{\vec{P}}_{11} \cdot [1, 1, ..., 1]
= \mbox{$\sum_{i=1}^N$} \, w_{\mu,i} \, f_i(\ten{F}; w_i^*)  \\[4.pt]
  \mu_{22}
= {\textsf{E}}[P_{22}] 
= \hat{\vec{P}}_{22} \cdot [1, 1, ..., 1]
= \mbox{$\sum_{i=1}^N$} \, w_{\mu,i} \, g_i(\ten{F}; w_i^*) 
\end{array}
\]
and the variances $\sigma_{11}$ and $\sigma_{22}$ are
\[
\begin{array}{@{\hspace*{1.8cm}}l}
  \sigma_{11}
={\textsf{Var}} [P_{11}] 
= \hat{\vec{P}}_{11} \cdot \ten{\varSigma} \cdot \hat{\vec{P}}_{11} \\[4.pt]
  \sigma_{22}
={\textsf{Var}} [P_{22}] 
= \hat{\vec{P}}_{22} \cdot \ten{\varSigma} \cdot \hat{\vec{P}}_{22}\,.
\end{array}
\]
For the special case with independent weights, 
$\ten{\varSigma} = \text{diag} \{ \sigma_1^2, \sigma_2^2, ..., \sigma_N^2 \}$, 
the variances simplify to
\[
\begin{array}{@{\hspace*{0.8cm}}l}
  \sigma_{11}
={\textsf{Var}} [P_{11}] 
= \mbox{$\sum_{i=1}^N$} \, (w_{\sigma,i} \, w_{\mu,i} \, f_i(\ten{F}; w_i^*))^2 \\[4.pt]
  \sigma_{22}
={\textsf{Var}} [P_{22}] 
= \mbox{$\sum_{i=1}^N$} \, (w_{\sigma,i} \, w_{\mu,i} \, g_i(\ten{F}; w_i^*))^2\,.
\end{array}
\]
This process allows us to compute the means and variances for the biaxial tensile stresses $P_{11}$ and $P_{22}$ for a sample given the model parameters. The stress means and variances are necessary  to compute the loss and to illustrate the model predictions. 
\begin{figure}[h]
\centering
\includegraphics[width=1.0\linewidth]{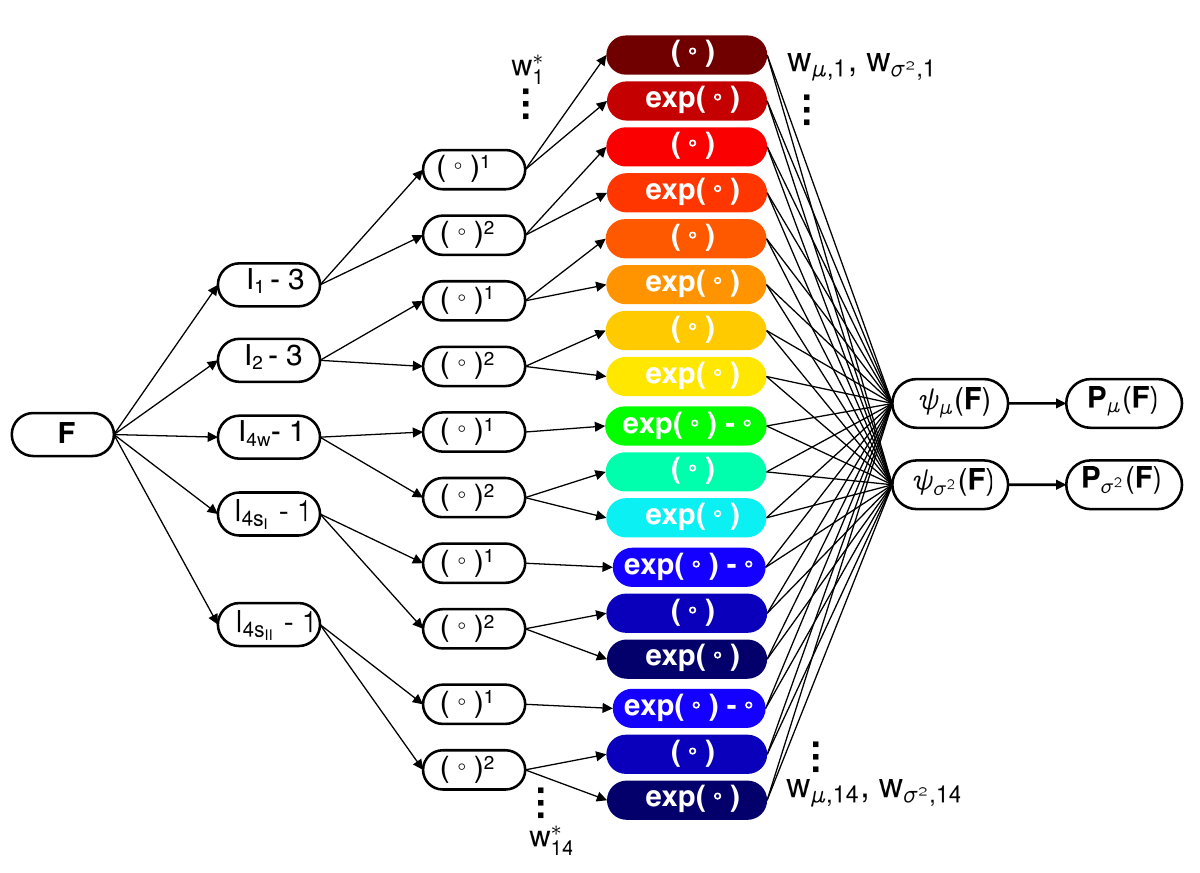}

\caption{{\bf{Gaussian constitutive neural network architecture.}} The first several layers of the network are identical to an ordinary constitutive neural network for anisotropic materials. The final layer has two sets of parameters, one which corresponds to the mean and one which corresponds to the variance. These terms are differentiated and summed together as summarized in Section \ref{sec:arch}.}
\label{arch}
\end{figure}
\subsection{Loss Function}
We want to find the maximum likelihood estimate of the model parameters $w_i^*$, $w_i^\mu$, and $\ten{\varSigma}$. To do this we must first compute the likelihood, which is the probability of observing our experimental data given some particular model parameters. Then, instead of {\it{maximizing the likelihood}}, we can equivalently {\it{minimize the negative log of the likelihood}} {\textsf{NLL}}. It is straightforward to compute the probability of observing a given value from a normal distribution, so we can write the negative log likelihood as 
$$
{\textsf{NLL}} 
= -\ln\prod_{i=1}^N \frac{1}{\sqrt{2 \pi \sigma_i^2}} 
\exp \left(-\frac{(P_i - \mu_i)^2}{2 \sigma_i^2} \right)\,, 
$$
which we can reformulate as follows,
$${\textsf{NLL}} 
= \frac{1}{2} N \ln(2 \pi \sigma_i^2) 
+ \frac{1}{2 \sigma_i^2} \sum_{i=1}^N (P_i - \mu_i)^2 \,.
$$
\subsection{Polyconvexity}
In Section \ref{sec:arch}, we have introduced the joint distribution of the weights $w_i$ as
$\hat{\vec{w}}
= [w_1/w_{\mu,1}, w_2 / w_{\mu,2}, \ldots, w_N / w_{\mu,N}] \sim {\mathcal{N}}([1, 1, \ldots, 1], \ten{\varSigma})$.
However, if the diagonal elements of $\ten{\varSigma}$ are very large, the probability that all weights remain positive, $w_i > 0$, can be as small as ${1}/{2}$. To ensure polyconvexity of the free energy function, 
we typically require that the internal and external weights are non-negative, $w_i^* \ge 0$ and $w_i \ge 0$. However, for any Gaussian random variable with non-zero variance, there is always a finite probability that the random variable will become negative. Thus, it is not feasible to require non-negativity with a probability of one. Instead, we will introduce a constraint that places an upper bound on the probability that a given weight is negative, 
${\textsf{p}}(w_i < 0)$. 
For each term $i$ with $w_i/w_{\mu,i} \sim N(1, \varSigma_{ii})$, 
this probability becomes, 
${\textsf{p}}(w_i < 0) = \varPhi(-1/\varSigma_{ii})$,
where $\varPhi$ denotes the cumulative distribution function for the standard normal distribution. Therefore, placing an upper bound on $\varSigma_{ii}$ places an upper bound on ${\textsf{p}}(w_i < 0)$. While there are many possible choices for an appropriate upper bound, we select $\varSigma_{ii} < 1$, which ensures that ${\textsf{p}}(w_i < 0) < \varPhi(-1) \approx 0.16$. This choice has the additional benefit that any stress that is one standard deviation below the mean, $\mu_i - \sigma_i$, remains polyconvex.
\subsection{Sparsity}\label{sec_sparsity}
By extending a constitutive neural network from deterministic to stochastic, inevitably, the number of parameters increases. A deterministic constitutive neural network with $N$ terms has $2N$ parameters. By contrast, a Gaussian constitutive neural network with $N$ terms has $3N$ parameters if we assume that the weights are independent, and $(N^2 + 5N)/2$ parameters if we assume that the weights are correlated. As a result, introducing a regularization penalty that promotes sparsity is particularly important to ensure the discovered model is interpretable. Using an $L_p$ regularization with $p=0.5$ on $w_{\mu,i}$ will promote sparsity \cite{mcculloch_sparse_2024}, because if $w_{\mu,i} = 0$ then $w_i = 0$ with a probability of one. Because of the polyconvexity constraint, $\varSigma_{ii} < 1$, if $w_{\mu,i} \to 0$, then $w_i$ converges in probability to 0 regardless of~$\ten{\varSigma}$.
For a deterministic constitutive neural network,
the expression for an $L_p$ regularization loss for $p=0.5$ is 
${\textit{L}}_{p} = \alpha \sum_{i = 1}^{N} (w_i \, f_i(w_i^*))^{0.5}$, where $f_i(w_i^*)$ is the integral of the contribution of the $i$-th term to the stress when $w_i = 1$. 
For a Gaussian constitutive neural network, 
we adjust the regularization loss as
$$
 {\textit{L}}_{p}
= \alpha \; \mbox{$\sum_{i = 1}^{N}$} \, (w_{\mu,i} \, f_i(w_i^*))^{0.5}
  \quad \text{for} \quad
  p=0.5 \,,
$$
and replace the external network weight $w_i$ by its mean $w_{\mu,i}$. In order to select the regularization parameter $\alpha$, we train with various values of $\alpha$ and chose the model with the fewest terms that does not significantly increase the test set negative log likelihood, where we define a significant increase as more than 0.1 above the minimum observed test set negative log likelihood. 
\subsection{Training, Development, and Test Sets}
In most machine learning applications, experimental data are divided into training, development, and test sets of different sizes by independently assigning each  example to one of the sets \cite{sohil_introduction_2021}. However, this is uncommon when the goal is to {\it{discover}} hyperelastic constitutive models \cite{linka_new_2023}. In our data, we have 10 distinct experiments, each with stress-stretch pairs in two directions \cite{mcculloch_automated_2024}. However, due to material symmetry, there are only 15 unique stress-stretch curves instead of 20. Furthermore, each stress-stretch curve contains 500 measurements, because there are five samples tested and 100 stress measurements are collected per experiment. Instead of randomly assigning each of the $n = 500 \cdot 15 = 7,500$ measurements into the training, development, or test set, we instead randomly assign three of the 15 stress-stretch curves to be the development set, while all remaining stress-stretch curves are used as the training set. This results in an 80/20 split between the training and development sets, which is common in other machine learning applications \cite{vabalas_machine_2019}. In particular, the development set included the s stress for the strip-w experiment and the w stress for the equibiaxial experiment in the 0/90 orientation, as well as the x stress for the strip-x experiment in the -45/+45 orientation.
This allows us to quantify the model's performance, not only on experiments it has seen before, but also on completely new loading configurations. Excluding three stress-stretch curves from the training data ensures that our model is actually discovering a generalizable model, and not is simply interpolating between data points.  

\section{Results}
\begin{figure}[h]
\centering
\includegraphics[width=1.0\linewidth]{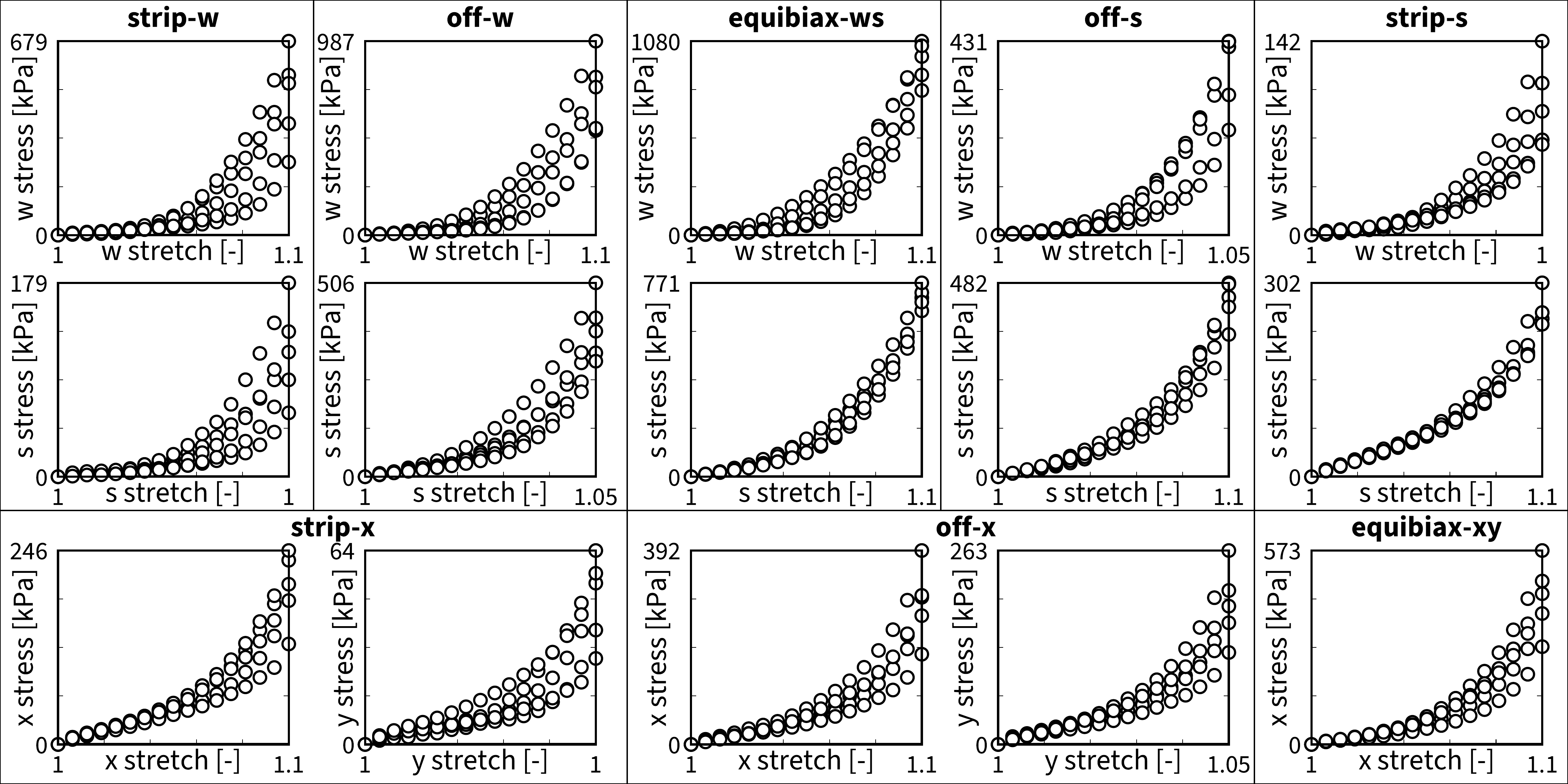}

\caption{{\bf{Experimental data from biaxial testing.}} The top two rows show experiments where the main or warp fiber direction is aligned with the loading direction, and the bottom row shows experiments where the main fiber direction is offset from the loading directions by 45 degrees. Each box represents a different experimental setup and shows data from all five samples.}
\label{raw_data}
\end{figure}

We collected the experimental data by loading Ethicon Prolene meshes using a Cellscale Biotester 5000, and can be seen in Figure \ref{raw_data}. We tested the samples in two different orientations, one with the main or warp fiber aligned with one of the loading directions, and one with the main fiber offset from the loading directions by 45 degrees. For each of these two orientations, we tested five samples, and loaded each sample with five different stretch ratios \cite{mcculloch_automated_2024}. Throughout each experiment, we recorded the forces and displacements from which we computed the stretch and stress. We collected $n=100$ data points per stretch ratio, and averaged the loading and unloading stresses to obtain a single value. 

\subsection{Unregularized, Independent Model}\label{sec_unregularized}
\begin{figure}[h]
\centering
\includegraphics[width=1.0\linewidth]{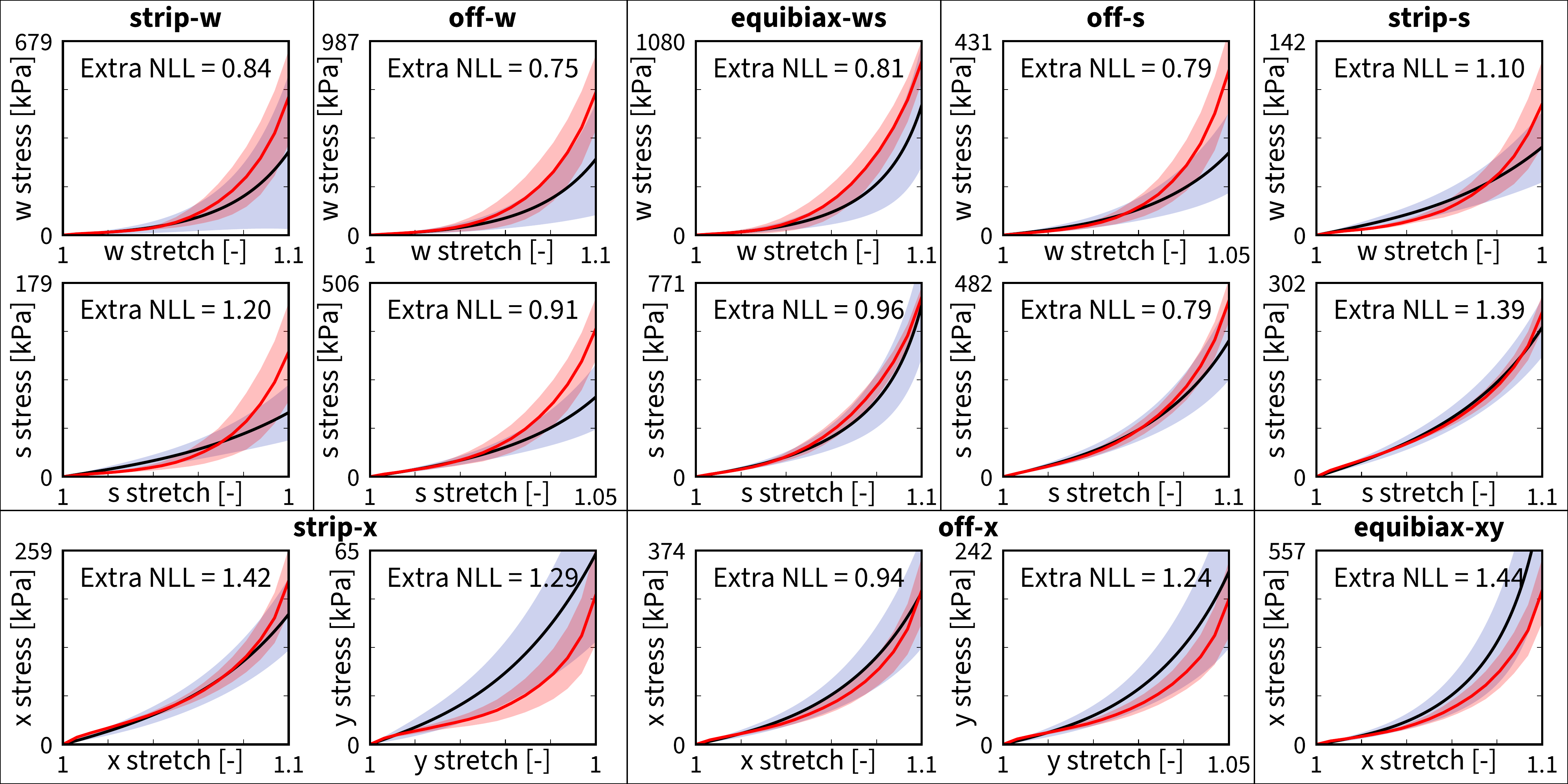}
\caption{{\bf{Experimental data and model prediction for unregularized, independent Gaussian constitutive neural network.}} The red line and shaded area show the mean and variance of the experimental data. The blue line and shaded area show the model predicted mean and variance. Within each plot, we report the extra {\textsf{NLL}} as the difference between the {\textsf{NLL}} and the minimum theoretical {\textsf{NLL}} given the data distribution, i.e., the {\textsf{NLL}} if the data and model distributions were identical.}
\label{unregularized}
\end{figure}

To start, we trained a Gaussian constitutive neural network, ignored the upper bound constraint on the variance, modeled the weights as independent, and set the regularization parameter $\alpha$ to zero. We initialized the model parameters randomly and used the ADAM optimizer to minimize the negative log likelihood loss. We used a learning rate of 0.001 and trained the model for 2,000 epochs with a batch size of 1,000. Figure \ref{unregularized} shows the resulting model predictions. The trained model has 12 non-zero terms, and achieves a training loss of 4.135 on the training set and 4.396 on the development set. Table \ref{losses_table} summarizes the performance of the unregularized model.

\begin{table}
\caption{{\bf{Performance of all models by the number of non-zero terms, loss on training set, and loss on development set.}}
The first block summarizes the performance of the unregularized, independent model from Section \ref{sec_unregularized}; the second block the performance of the regularized, independent model from Section \ref{sec_independent}; and the third block the regularized, correlated model from Section \ref{sec_correlated}. Grey colors highlight the model with the fewest terms that does not significantly decrease negative log likelihood as defined in section \ref{sec_sparsity}, associated with Figures \ref{unregularized}, \ref{independent}, and \ref{correlated}.}

\vspace*{0.2cm}
\begin{tabular}{|lcccc|}
\hline
\textbf{model} & \textbf{$\alpha$} & {\!\!\!\textbf{\# terms}\!\!} & {\!\textbf{train \textsf{NLL}}\!} & \textbf{dev \textsf{NLL}} \\
 \hline \hline
\rowcolor{gray} unconstrained & 0.00 & 12 & 4.135     & 4.396    \\ \hline \hline
independent   & 0.00  & 12  & 4.133     & 4.427    \\
independent   & 0.01  & 9   & 4.132     & 4.426    \\
independent   & 0.03  & 6   & 4.143     & 4.428    \\
\rowcolor{gray} independent   & 0.10  & 5   & 4.134     & 4.452    \\
independent   & 0.30  & 6   & 4.163     & 4.503    \\
independent   & 1.00  & 3   & 4.249     & 4.577    \\ \hline \hline
correlated    & 0.00  & 10  & 4.076     & 4.306    \\
correlated    & 0.01  & 8   & 4.077     & 4.305    \\
correlated    & 0.03  & 8   & 4.078     & 4.313    \\
\rowcolor{gray} correlated    & 0.10  & 4   & 4.085     & 4.340    \\
correlated    & 0.30  & 4   & 4.116     & 4.343    \\
correlated    & 1.00  & 3   & 4.325     & 4.630    \\\hline
\end{tabular}\newline
\label{losses_table}
\end{table}

\subsection{Regularized, Independent Model}\label{sec_independent}
\begin{figure}[h]
\centering
\includegraphics[width=1.0\linewidth]{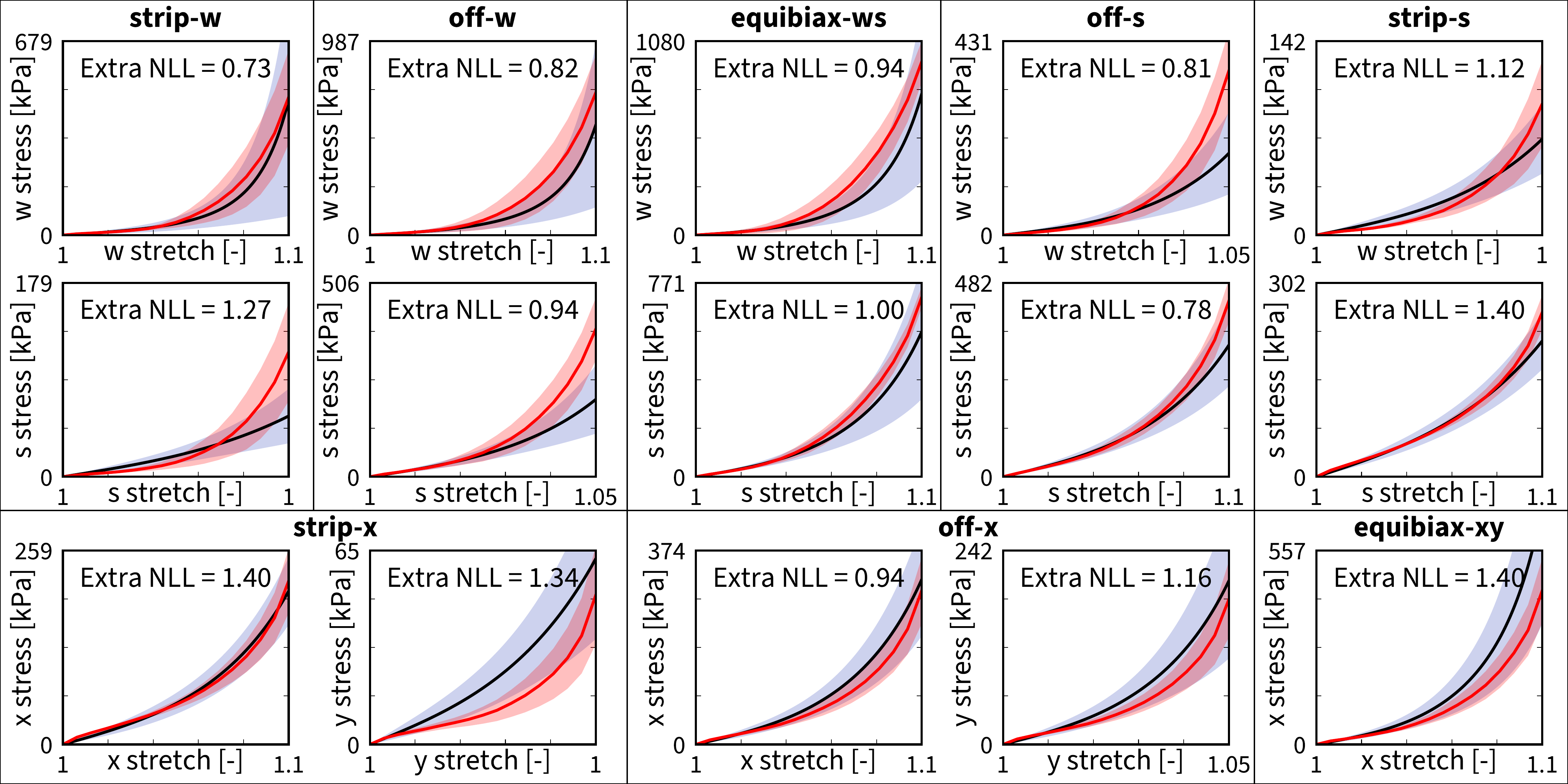}
\caption{{\bf{Experimental data and model prediction for regularized, independent Gaussian constitutive neural network with $\alpha = 0.10$}}. The red line and shaded area show the mean and variance of the experimental data. The blue line and shaded area show the model predicted mean and variance. Within each plot, we report the extra {\textsf{NLL}} as the difference between the {\textsf{NLL}} and the minimum theoretical {\textsf{NLL}} given the data distribution, i.e., the {\textsf{NLL}} if the data and model distributions were identical.}
\label{independent}
\end{figure}

Next, we trained a Gaussian constitutive neural network where we modeled the weights as independent, constrained the variance such that $\varSigma_{ii} < 1$, and set the regularization parameter $\alpha$ to values between 0.0 and 1.0. We initialized the model parameters randomly and used the ADAM optimizer to minimize the {\textsf{NLL}} loss. We used a learning rate of 0.001 and trained the model with a batch size of 1,000. We pre-trained for 2,000 epochs with $\alpha = 0$, then without reinitializing the model parameters, trained for 2,000 more epochs with non-zero $\alpha$. Figure \ref{independent} shows the model predictions for $\alpha = 0.1$, which we selected as described in section \ref{sec_sparsity}. This model has five non-zero terms, and achieves a training loss of 4.134 on the training set and 4.452 on the development set.
Its strain energy is
\[
\begin{array}{l@{\hspace*{0.1cm}}c@{\hspace*{0.1cm}}l@{\hspace*{0.1cm}}c@{\hspace*{0.1cm}}l}
\psi &=& \frac{1}{2}a_1(-1 + \exp(b_1(I_1 - 3))) / b_1\\[3.pt]
&+&\frac{1}{2}a_2(-1 + \exp(b_2(I_2 - 3)^2)) / b_2 \\[3.pt]
&+&\frac{1}{2}a_3(-1 + \exp(b_3(I_{4w} - 1)^2)) / b_3\\[3.pt]
&+&\frac{1}{2}\mu_1(I_{4s} - 1)^2  \\[3.pt]
&+&\frac{1}{2}a_4(-1 + \exp(b_4(I_{4s_I} - 1)^2)) / b_4 \\[3.pt]
&+&\frac{1}{2}a_4(-1 + \exp(b_4(I_{4s_{II}} - 1)^2)) / b_4  \,,
\end{array} 
\]
where $a_1 \,{\sim}\, {\mathcal{N}}(40\text{ kPa},$ $ (40\text{ kPa})^2)$, $b_1 = 0.0057$, $a_2 \,{\sim}\,N(497\text{ kPa,}$ $ (497\text{ kPa})^2)$, $b_2 = 24$, $a_3 \,{\sim}\, {\mathcal{N}}(20\text{ kPa}, $ $(20\text{ kPa})^2)$, 
$b_3 = 69$, $\mu_1 \,{\sim}\, {\mathcal{N}}(39\text{ kPa},$ $ (0\text{ kPa})^2)$, $a_4 \,{\sim}\, {\mathcal{N}}(74\text{ kPa},$ and $ (0\text{ kPa})^2)$, $b_4 = 11$.
Table \ref{losses_table} summarizes the performance of the independent models for all regularization parameters $\alpha$.

\subsection{Regularized, Correlated Model}\label{sec_correlated}
\begin{figure}[h]
\centering
\includegraphics[width=1.0\linewidth]{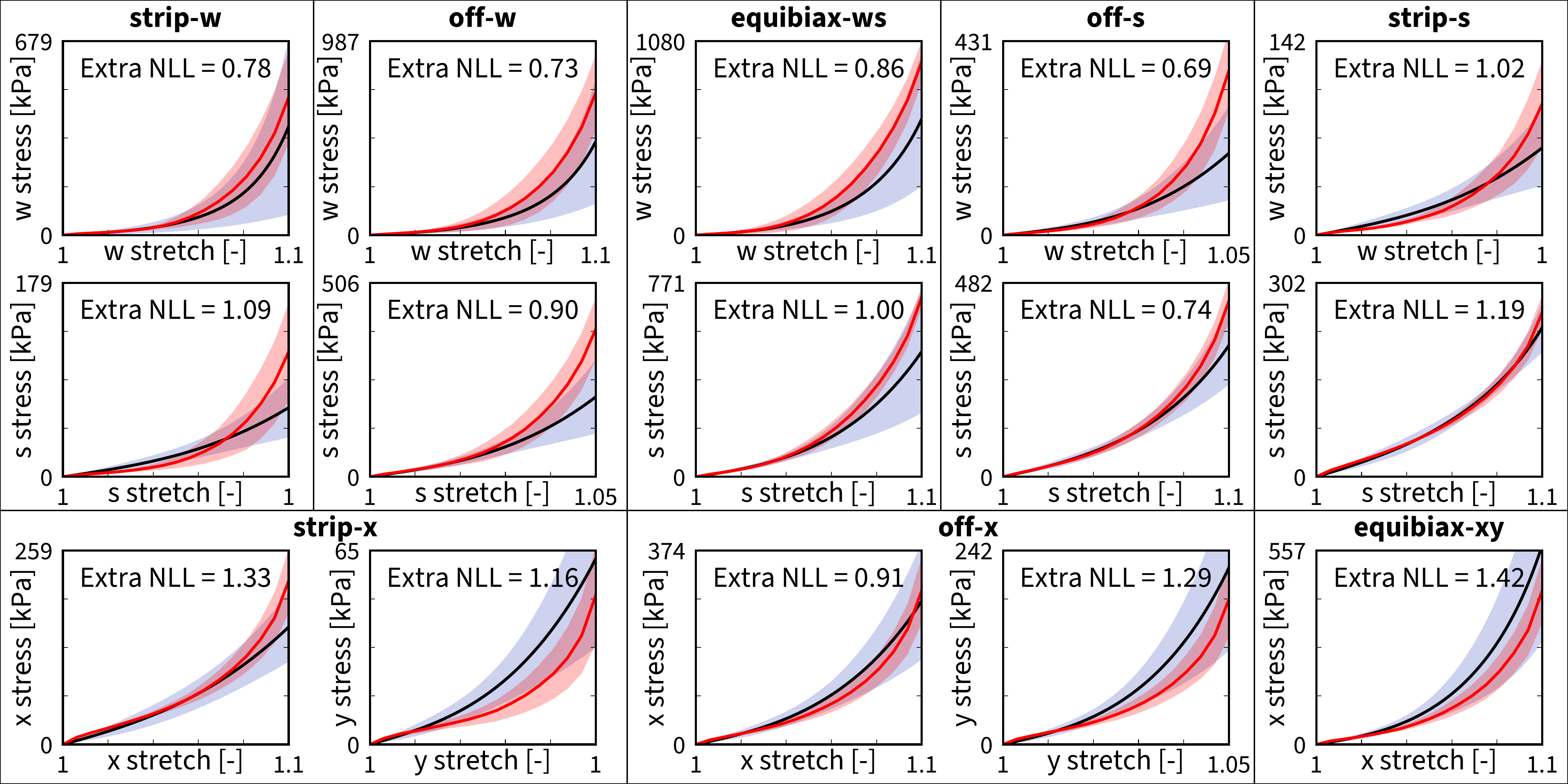}
\caption{{\bf{Experimental data and model prediction for the regularized, correlated Gaussian constitutive neural network with $\alpha = 0.10$.}} The predicted s-stress in the strip-s and off-s experiments has a smaller variance, and thus matches the data distribution more closely than the previous models. The red line and shaded area show the mean and variance of the experimental data. The blue line and shaded area show the model predicted mean and variance. Within each plot, we report the extra {\textsf{NLL}} as the difference between the {\textsf{NLL}} and the minimum theoretical {\textsf{NLL}} given the data distribution, i.e., the {\textsf{NLL}} if the data and model distributions were identical.}
\label{correlated}
\end{figure}

Finally, we trained a Gaussian constitutive neural network where we still enforced the constraint that $\varSigma_{ii} < 1$ and set the regularization parameter $\alpha$ to values ranging from 0.0 to 1.0, but did not constrain the covariance matrix for the weights to be diagonal. We initialized the model parameters randomly and used the ADAM optimizer to minimize the negative log likelihood loss. We used a learning rate of 0.001 and trained the model with a batch size of 1,000. We pre-trained for 2,000 epochs with $\alpha = 0$, then without reinitializing the model parameters, trained for 2,000 more epochs with non-zero $\alpha$.  Figure \ref{correlated} shows the model predictions for $\alpha = 0.1$, which we chose as described in section \ref{sec_sparsity}. This model has four non-zero terms, and achieves a training loss of 4.085 on the training set and 4.340 on the development set. The discovered strain energy is 
\[
\begin{array}{l@{\hspace*{0.1cm}}c@{\hspace*{0.1cm}}l@{\hspace*{0.1cm}}c@{\hspace*{0.1cm}}l}
\psi 
&=& \frac{1}{2}a_1(-1 + \exp(b_1(I_{4w} - 1)^2)) / b_1 + \frac{1}{2}\mu_1(I_2 - 3)^2 \\[3.pt]
&+&\frac{1}{2}a_2(-1 + \exp(b_2(I_{4s} - 1)^2)) / b_2 \\[3.pt]
&+&\frac{1}{2}a_3(-b_3(I_{4s_I}\; - 1) - 1 + \exp(b_3(I_{4s_I} \;- 1))) / b_3 \\[3.pt]
&+&\frac{1}{2}a_3(-b_3(I_{4s_{II}} - 1) - 1 + \exp(b_3(I_{4s_{II}} - 1))) / b_3 \\[3.pt]
\end{array}
\]
where $\mu_1 \,{\sim}\, {\mathcal{N}}(777\text{ kPa},$ $ (777\text{ kPa})^2)$, 
$a_1 \,{\sim}\, {\mathcal{N}}(38\text{ kPa},$ $ (38\text{ kPa})^2)$, $b_1 = 49$, $a_2 \,{\sim}\,N(19\text{ kPa,}$ $ (19\text{ kPa})^2)$, 
$b_2 = 33$, 
$a_3 \,{\sim}\, {\mathcal{N}}(26278\text{ kPa}, $ $(12783\text{ kPa})^2)$, and
$b_3 = 0.0083$, 
and the correlation matrix for the weights $[\, \mu_1, a_1, a_2, a_3 \,]$ is 
$$
{\textsf{Corr}}
= \begin{bmatrix}
+1.00 & -0.23 & -0.10 & -0.50\\
-0.23 & +1.00 & +0.05 & +0.32\\
-0.10 & +0.05 & +1.00 & -0.78\\
-0.50 & +0.32 & -0.78 & +1.00\\
\end{bmatrix}\,.$$
Table \ref{losses_table} summarizes the performance of the correlated models for all regularization parameters $\alpha$.
\section{Discussion}\label{discussion}
We developed a Gaussian constitutive neural network that discovers stochastic constitutive models to predict probability distributions of the stress as functions of the deformation gradient while satisfying physical constraints. Compared to existing Bayesian constitutive neural networks \cite{linka_discovering_2025}, these {\it{Gaussian constitutive neural networks do not require prior selection}}. Other stochastic automated constitutive model discovery methods that are built on bayesian neural networks similarly require prior selection, and often use either gaussian or spike and slab priors \cite{joshi_bayesian-euclid_2022,olivier_bayesian_2021}. By choosing the hyperparameters of these priors, it is thus inevitable that with a bayesian neural network the discovered weights and by extension the model prediction will be affected by the arbirtrary choice of prior. 

Using a comprehensive data set from biaxial experiments, we compared three classes of Gaussian constitutive neural networks: unregularized indepenent, regularized independent, and regularized correlated. Figures \ref{unregularized}, \ref{independent}, \ref{correlated} and Table \ref{losses_table} demonstrate that our Gaussian constitutive neural network successfully discovered models and parameters for all three classes. Notably, our model is verifiable, as we can see visually and by computing the extra negative log likelihood that the model accurately predicts the mean and variance of the distribution. Other stochastic and generative constitutive model discovery methods separately fit a model for each sample and then attempt to characterize the distribution of the weights \cite{tac_generative_2024}, which does not directly provide any information about the probability distribution of the stress. The ability to directly predict the mean and variance of the stress is also useful when applying constitutive models in simulations as it provides error bounds on the predictions of the simulation. 

Our results suggest that {\it{the correlated model outperforms the independent model}}. As we can see in Table \ref{losses_table}, the correlated model has a smaller loss than the independent model, on both the training set and the test set. From comparing Figures \ref{independent} and \ref{correlated}, we conclude that the correlated model achieves a better fit for the s-stress of the strip-s experiment; consequently, the extra {\textsf{NLL}} for this plot decreased from 1.40 to 1.19. In particular, the independent model predicts a stress variance that is much larger than the variance of the data, while the correlated model predicts a stress variance that is closer to the variance of the data. In the independent model, the loss is dominated by the experiments where the data have a large stress variance because the negative log likelihood more strongly penalizes underestimating the model variance than it penalizes overestimating the model variance. As a result, the independent model accurately characterizes the variance for experiments with large variance while overestimating the variance for experiments with a small variance.

For the correlated model, the fact that {\it{all weights can be positively or negatively correlated}} makes it possible for the model to predict large  variances for some experiments and small  variances for others. Consequently, the performance of the model on experiments with a large  variance does not change significantly, while the performance of the model on experiments with a small  variance improves significantly. Taken together, our observations suggest that allowing the weights to be correlated results in a better agreement between the predicted stress variance and the variance of the measured stresses across experiments. 

Can we use Gaussian constitutive neural networks in generative machine learning?
Generative machine learning models enable us to model and sample from complex distributions such as the distribution of images or texts. When applied to constitutive modeling, generative machine learning would allow us to learn the distribution of material parameters and thus allows us to sample a prototypical constitutive model from our discovered distribution. Generative models in constitutive modeling can also be used to learn constitutive models for new material samples given limited data \cite{yin_generative_2023}. {\it{Gaussian constitutive neural networks are generative machine learning models}} which assume that each material parameter is distributed according to a normal distribution. When training with small amounts of data, this simple model may be sufficient to fit the data distribution; however, when training with large amounts of data, it may become clear that the model weights are not distributed according to a normal distribution. 

When the model weights are not normally distributed, a Gaussian constitutive neural network can still be a useful component in a more complex generative model to ensure that physical constraints are followed. Most generative models including variational autoencoders and generative adversarial networks include a neural network that computes the model output given the latent variables and the model parameters \cite{yin_generative_2023}. Rather than using a fully-connected neural network to compute the stress distribution as a function of the deformation gradient and the latent variables, we could use a fully-connected neural network to compute the Gaussian constitutive neural network parameters as a function of the latent variables, and then use a Gaussian constitutive neural network to compute the stress mean and variance as a function of the deformation gradient and the output of the fully-connected network. With this modification, the model can still be trained by using gradient descent to maximize the evidence lower bound. This would enable the design of increasing complex generative models that exactly satisfy physical constraints. 

\section{Conclusions}
By integrating constitutive neural networks and Gaussian neural networks, we have created a stochastic neural network model, which is motivated by physical principles and accounts for variations between samples. By using an anisotropic constitutive neural network as a template, we ensure that our model satisfies known physical constraints such as material symmetry and polyconvexity by design. Furthermore, by treating some weights in the neural network as random variables, we can predict not only the mean stress across all samples, but also the probability distribution of the stresses across all samples. Previous work addressed stochastic effects using Bayesian constitutive neural networks that suffer from a lack of interpretability, a need to choose priors for all model weights, and non-deterministic model outputs. Here we developed a method, which models each stochastic network weight with only two scalar parameters, and analytically computes the probability distribution of the stress. We designed three classes of these Gaussian constitutive neural networks: unregularized indepenent, regularized independent, and regularized correlated. The latter
models the weights as stochastic, distributed according to a multivariate Gaussian distribution, and allows the weights to be correlated. We all three classes of models on data from biaxial tension tests. For the independent class, we discovered a five-term model that fits the training and development sets well. For the correlated class, we discovered a
four-term model that was more accurate on both the training and development sets, but had more non-zero parameters and was thus less interpretable. Our Gaussian constitutive neural networks can be used to discover models and distributions of weights across a set of samples, which can be used as priors for discovering an accurate constitutive model for a new sample with limited data. We anticipate that Gaussian constitutive neural networks are a valuable first step towards building generative machine learning models such as variational autoencoders to generate constitutive models for multimodal data from different classes of materials. 
\section*{Data availability}
\noindent
Our source code, data, and examples are available at \\
https://github.com/LivingMatterLab/CANN.
\section*{Acknowledgments}
\noindent
This work was supported by 
an NSF Graduate Student Fellowship to Jeremy McCulloch, and
the ERC Advanced Grant 101141626 and the NSF CMMI grant 2320933 to Ellen Kuhl. \\[4.pt]

\end{document}